\documentclass[twocolumn]{aastex61}
\pdfoutput=1 %for arXiv submission
\usepackage{amsmath,amstext}
\usepackage[T1]{fontenc}
\usepackage{apjfonts} 
\usepackage[figure,figure*]{hypcap}
\usepackage{mypack}

\shorttitle{A Massive-born NS with a Massive WD Companion}
\shortauthors{Cognard, Freire, Guillemot et al.}

\begin{document}

\title{A Massive-born Neutron Star with a Massive White Dwarf Companion}

\author{Isma\"el Cognard}
\affiliation{Laboratoire de Physique et Chimie de l'Environnement et de l'Espace, Universit\'e d'Orl\'eans/CNRS, F-45071 Orl\'eans Cedex 02, France}
\affiliation{Station de radioastronomie de Nan\c{c}ay, Observatoire de Paris, CNRS/INSU, F-18330 Nan\c{c}ay, France}
\author{Paulo C. C. Freire}
\affiliation{Max-Planck-Institut f\"{u}r Radioastronomie, Auf dem H\"{u}gel 69, D-53121 Bonn, Germany}
\affiliation{Station de radioastronomie de Nan\c{c}ay, Observatoire de Paris, CNRS/INSU, F-18330 Nan\c{c}ay, France}
\author{Lucas Guillemot}
\affiliation{Laboratoire de Physique et Chimie de l'Environnement et de l'Espace, Universit\'e d'Orl\'eans/CNRS, F-45071 Orl\'eans Cedex 02, France}
\affiliation{Station de radioastronomie de Nan\c{c}ay, Observatoire de Paris, CNRS/INSU, F-18330 Nan\c{c}ay, France}
\author{Gilles Theureau}
\affiliation{Laboratoire de Physique et Chimie de l'Environnement et de l'Espace, Universit\'e d'Orl\'eans/CNRS, F-45071 Orl\'eans Cedex 02, France}
\affiliation{Station de radioastronomie de Nan\c{c}ay, Observatoire de Paris, CNRS/INSU, F-18330 Nan\c{c}ay, France}
\affiliation{LUTH, Observatoire de Paris, PSL Research University, CNRS, Universit\'e Paris Diderot, Sorbonne Paris Cit\'e, F-92195 Meudon, France}
\author{Thomas~M. Tauris}
\affiliation{Max-Planck-Institut f\"{u}r Radioastronomie, Auf dem H\"{u}gel 69, D-53121 Bonn, Germany}
\affiliation{Argelander-Institut f\"ur Astronomie, Universit\"at Bonn, Auf dem H\"ugel 71, D-53121 Bonn, Germany}
\author{Norbert Wex}
\affiliation{Max-Planck-Institut f\"{u}r Radioastronomie, Auf dem H\"{u}gel 69, D-53121 Bonn, Germany}
\author{Eleni Graikou}
\affiliation{Max-Planck-Institut f\"{u}r Radioastronomie, Auf dem H\"{u}gel 69, D-53121 Bonn, Germany}
\author{Michael Kramer}
\affiliation{Max-Planck-Institut f\"{u}r Radioastronomie, Auf dem H\"{u}gel 69, D-53121 Bonn, Germany}
\author{Benjamin Stappers}
\affiliation{Jodrell Bank Center for Astrophysics, School of Physics and Astronomy, The University of Manchester, M13 9PL, UK}
\author{Andrew G. Lyne}
\affiliation{Jodrell Bank Center for Astrophysics, School of Physics and Astronomy, The University of Manchester, M13 9PL, UK}
\author{Cees Bassa}
\affiliation{ASTRON, The Netherlands Institute for Radioastronomy, Postbus 2, 7900 AA, Dwingeloo, The Netherlands}
\author{Gregory Desvignes}
\affiliation{Max-Planck-Institut f\"{u}r Radioastronomie, Auf dem H\"{u}gel 69, D-53121 Bonn, Germany}
\author{Patrick Lazarus}
\affiliation{Max-Planck-Institut f\"{u}r Radioastronomie, Auf dem H\"{u}gel 69, D-53121 Bonn, Germany}

%=========================================================================================

\begin{abstract} We report on the results of a 4-year timing campaign of
PSR~J2222$-0137$, a 2.44-day binary pulsar with a massive white dwarf (WD)
companion, with the Nan\c{c}ay, Effelsberg and Lovell radio telescopes. 
Using the Shapiro delay for this system, we find a pulsar mass
$m_{p} \, = \, 1.76\,\pm\,0.06\,\Msun$ and a WD mass  $m_{c} \, = \, 1.293\,\pm\,0.025\,\Msun$.
We also measure the rate of advance of periastron for this system, which
is marginally consistent with the GR prediction for these masses.
The short lifetime of the massive WD progenitor star led to a rapid X-ray binary
phase with little ($< \, 10^{-2} \, \Msun$) mass accretion onto the neutron star (NS); hence,
the current pulsar mass is, within uncertainties, its birth mass,
which is the largest measured to date.
We discuss the discrepancy with previous mass measurements for this
system; we conclude that the measurements presented here are likely to be more accurate.
Finally, we highlight the usefulness of this system for testing alternative theories of gravity
by tightly constraining the presence of dipolar radiation. This is of particular importance 
for certain aspects of strong-field gravity, like spontaneous scalarization, since 
the mass of PSR~J2222$-0137$ puts that system into a poorly tested parameter range.
\end{abstract}

\keywords{pulsars: general --- pulsars: individual J2222$-$0137 --- stars: neutron star 
          --- white dwarfs --- binaries: close --- X-rays: binaries}

%\maketitle

%=========================================================================================

\section{Introduction}
\label{sec:introduction}

%------------------------------------------------------------------------------------------

\subsection{The PSR~J2222$-$0137 binary system}

PSR~J2222$-$0137 is a recycled pulsar ($P\, = \, 32.8\, \rm ms$,
$B\,=\,0.76\, \times\, 10^9\,$G) discovered in
the Green Bank Telescope (GBT) 350 MHz drift scan survey \citep{Boyles}. It is in a binary system with
orbital period $P_b\, = \,2.44576\, \rm d$, the projected semi-major axis of the pulsar's
orbit is $x\, = \,$10.848 light seconds (lt-s).

This results in a mass function of $0.229 \, \Msun$, which
implies a relatively massive companion: in a follow-up paper,
Kaplan et al. (2014, henceforth Paper I)
estimated the masses of the components of the system to be
$m_p \,=\,1.20 \, \pm 0.14\, \Msun$ for the pulsar,
$m_c \, = \,1.05 \, \pm 0.06 \, \Msun$ for the companion and an orbital inclination of
$i \, = \, 86.8\, \pm \, 0.04^\circ$ (or $i' \, = 180^\circ -i \, = \, 93.2\, \pm \, 0.04^\circ$),
from the Shapiro delay in this system. As discussed
in Paper I, the orbital eccentricity of the system
($e\, = \,0.00038$) implies that the companion is a white dwarf (WD) star:
the formation of a second neutron star (NS) would have induced an
orbital eccentricity (from kicks and, even in the case of a symmetric explosion,
from the sudden loss of the binding energy of
the NS) at least two orders of magnitude larger,
with no possibility of subsequent tidal circularization of the
orbit. The system can therefore be classified as an intermediate mass binary pulsar
(IMBP, see \citealt{Camilo96,TLK12}).
Although small, the reported pulsar mass was not considered surprising as
it is similar to that of another IMBP, PSR~J1802$-$2124, with
$m_p \,=\,1.24 \, \pm \, 0.11 \, \Msun$ and $m_c \,=\,0.78 \, \pm \, 0.04 \, \Msun$
\citep{Ferdman2010}.

PSR~J2222$-$0137 has a dispersion measure (DM) of only 3.28 cm$^{-3}$pc,
one of the lowest for any pulsar. According to the NE2001 model \citep{CordesLazio2001} of
the electron distribution in the Galaxy, this implies a distance of 312 pc.
This motivated a Very Long Baseline Interferometry (VLBI)
astrometric campaign that obtained the most precise
VLBI distance for any pulsar, $d \, = \, 267.3^{+1.2}_{-0.9}$ pc (\citealt{Deller}, henceforth
Paper~II); 15 \% smaller than the NE2001 prediction. They also measured 
unusually precise values for the position and proper motion of the system: the total proper 
motion is large ($\mu_T\,=\,45.09(2)$ mas yr$^{-1}$), but the implied transverse velocity
($v_{T} \, = \, 57.1^{+0.3}_{-0.2} \, \rm km \, s^{-1}$ in the Barycentric reference frame)
is typical among recycled pulsar systems (e.g., \citealt{Gonzalez2011}).
Because of the relatively large size of the pulsar orbit and
the proximity of the system, \cite{Deller} were able to detect, for the first time,
the orbital motion of the pulsar in the astrometric data; from this they
derived a position angle (PA) of the line of nodes,
$\Omega \, = \, 5^{+15}_{-20}$$^\circ$.

The relatively small distance to this binary pulsar allowed
detailed multi-wavelength follow-up.
\cite{Kaplan} observed the astrometric position of
PSR~J2222$-$0137 at optical wavelengths with the Keck I and II telescopes, and surprisingly the WD
companion was not detected; this implies that it is at least 100 times fainter
than any other WD companion to a pulsar detected to date. Whether this is
surprising or not depends on the temperature and radius of this WD. 
Assuming a WD mass of $1.05\, \Msun$, a WD radius of about
0.8 Earth radii ($R_{\oplus}$) was derived in Paper I. For this radius,
the optical non-detection implies a surface temperature smaller than 3000~K. According
to the WD cooling models of \cite{Bergeron11} such a low temperature is
only reached after the rapid cooling phase sets in following crystalization of
the core, and the estimated cooling age is about 8~to~10~Gyr. The characteristic
age of the pulsar is about 30 Gyr, which implies that an age of 8 - 10 Gyr
cannot be excluded from spin considerations\footnote
{Although the characteristic age does not provide a reliable age estimate
for a recycled pulsar \citep{TLK12}, it represents an approximate upper
limit for that age, assuming a braking index of 3.
Therefore, if the characteristic age of PSR~J2222$-$0137 were
much smaller than 8 Gyr, we would be able to exclude the
extreme age suggested by the cooling timescale.
}.

%------------------------------------------------------------------------------------------

\subsection{Motivation and outline of this work}

We report on the results of a 4-year timing campaign of this
pulsar with the Nan\c{c}ay (NRT), Effelsberg and Lovell radio telescopes.
The motivation is to obtain more precise physical parameters for this system,
in particular the mass of PSR~J2222$-$0137, but also the mass of its WD companion.

The short stellar lifetime of the massive WD progenitor led to a rapid X-ray
binary phase resulting, as discussed below, in little
mass accretion onto the pulsar. Therefore, its current mass must be very similar to
its birth mass. This contrasts with the evolution of the fastest-spinning pulsars
which generally have much lower-mass companions; in those cases the long accretion
phases that spun them up could in principle have substantially increased the NS mass.

Precise NS birth masses are important because they probe the final 
stellar evolution phases of massive stars as well as supernova explosion
physics \citep{Ozel2012}. Most NS birth masses (and certainly the most precise among
them) have been measured in double neutron star systems (DNSs). Until 2013
all NSs in DNS systems had masses between 1.23 and 1.44$\,\Msun$; this situation
changed with the discovery of PSR~J0453+1559 \citep{Deneva2013}: for this
system $m_p \, = \, 1.559 \, \pm \, 0.005\, \Msun$ and
$m_c \, = \, 1.174 \, \pm \, 0.004\, \Msun$ \citep{Martinez2015}. This implies
that NSs are born with a wider range of masses than previously thought.
This highlights the importance of increasing the sample of NS birth masses:
we still don't know how wide the distribution is and what the maximum is,
and whether it is a single distribution,
bimodal or more complex \citep{Ozel&Freire2016,Antoniadis2017}.
Furthermore, increasing the (very small)
sample of NS birth masses in systems with massive WD companions has the added
benefit of verifying whether the companion mass and evolution have any effect
on this distribution.

The mass measurements in Paper I do not usefully place PSR~J2222$-$0137
within the observed NS birth mass range: within 2
$\sigma$, its mass is consistent with 0.92 and 1.48$\,\Msun$.
This motivated the long-term timing reported here that has yielded
improved (and significantly larger)
masses mainly by improving the measurement of the Shapiro delay and separating
it from the rate of advance of periastron.

The remainder of the paper is structured as follows.
In Section~\ref{sec:observations} we describe the details of the observations,
data processing, and present two timing solutions for the pulsar
with a description of how they were derived; one with positional fitting,
the other using the previously derived VLBI position.
The reasons for this are discussed in Section~\ref{sec:astrometry},
where we make a detailed comparison of the astrometry derived
from the timing with the VLBI astrometry.
In Section~\ref{sec:results} we present further timing results:
an improved measurement of the Shapiro delay (we discuss the resultant
component masses), a highly significant measurement of the rate
of advance of periastron $\dot{\omega}$ and finally discuss
the kinematic effects on the timing parameters.
In Section~\ref{sec:GWs}, we discuss the measurement of
the variation of the orbital period
($\dot{P}_{\rm b}$) and the resulting limits on the emission of
dipolar gravitational waves, we also discuss how precise this
system might become with future timing.
In Section~\ref{sec:modelling}, we discuss the past history of this system
in light of the new mass measurements. Finally, in Section~\ref{sec:conclusions},
we summarize our findings and discuss the prospects for continued timing
of this system.

%==========================================================================================

\section{Observations and data processing}
\label{sec:observations}

\begin{table*}
  \caption{Observations of J2222$-$0137 and data reduction parameters}
  \begin{center}
  \begin{tabular}{lcccc}
  \hline
  Telescope & Nan\c{c}ay L & Nan\c{c}ay S & Effelsberg & Lovell \\
  \hline
  Start of observations (MJD) \dotfill  & 56191 & 56204 &  57321 & 56251 \\
  End of observations (MJD)   \dotfill & 57527 & 57574 & 57766 & 57543 \\
  Bandwidth (MHz) \dotfill & 512 & 512 & 150/250 & 400 \\
  Bandwidth per TOA (MHz) \dotfill & 128 & 128 & 150/125 & 80 \\
  Center frequency (MHz) & 1484 & 2539 & 1360 & 1532 \\
  Number of TOAs used in solution \dotfill & 1601 & 80 & 939 & 258 \\
  Time per TOA (s) \dotfill & 600 & 600 & 600 & 600 \\
  Weighted residual rms ($\mu$s) \dotfill & 4.1 & 7.3 & 2.5 & 8.2 \\
  EFAC \dotfill & 1.0 & 1.0 & 1.0 & 1.0 \\
  EQUAD ($\mu$s) \dotfill & 2.46 & 0.0 & 0.96 & 4.46 \\
  \hline
 \end{tabular}
 \end{center}
 \label{table:observations}
\end{table*}

Our observations of PSR~J2222$-$0137 are summarized in Table~\ref{table:observations}.
The pulsar has been observed regularly with the ``low'' (1.2 -- 1.7\,GHz, or ``L band'')
and ``high'' (1.7 -- 3.0\,GHz, or ``S band'') receivers of the NRT for the $\sim$1 hour
the pulsar stays near transit since 2012 September 21; the last observation
used in this work was on 2016 July 5. We used the Nan\c{c}ay Ultimate
Pulsar Processing Instrument (NUPPI), a back-end similar to the Green Bank
Ultimate Pulsar Processing Instrument (GUPPI) at the GBT\footnote{
http://safe.nrao.edu/wiki/bin/view/CICADA/GUPPISupportGuide} to process the signal
which allows for real-time coherent dedispersion and folding of a bandwidth BW =
512 MHz \citep{Liu}. We note that the NUPPI data analyses presented in
\citet{Liu} were affected by a lack of precision while reading crucial internal
MJD dates by the {\tt PSRCHIVE} \citep{PSRCHIVE,PSRCHIVE2012} software, causing
reduced timing precision. This issue
has now been fixed and the corrected software has been extensively tested using
NUPPI datasets.

From 2015 October 26 to November 28, and again between 2017 January 5 to 13,
we conducted two intensive 
campaigns with the Effelsberg 100-m radio telescope, with the aim of confirming
(and possibly improving) the Shapiro delay measurements. Those data were taken
with the 20-cm receiver (with BW = 150 MHz) and the central beam of the 7-beam
system (BW = 250 MHz) using PSRIX \citep{PSRIX} as a back-end.
All observing data is presented in Table~\ref{table:observations}.

Observations of PSR~J2222-0137 were made with the 76-m Lovell Telescope at Jodrell Bank using a 400-MHz total bandwidth, starting 2012 November 20 until 2016 June 4.  The 400-MHz band was centred on 1532 MHz and was analysed using roach-based coherent dedispersion processing across five 80-MHz sub-bands.  Each observation had a duration of between six and twenty-five minutes.

\subsection{Polarimetry}
\label{sec:polarimetry}

\begin{figure*}
\centering
\includegraphics[width=0.8\textwidth]{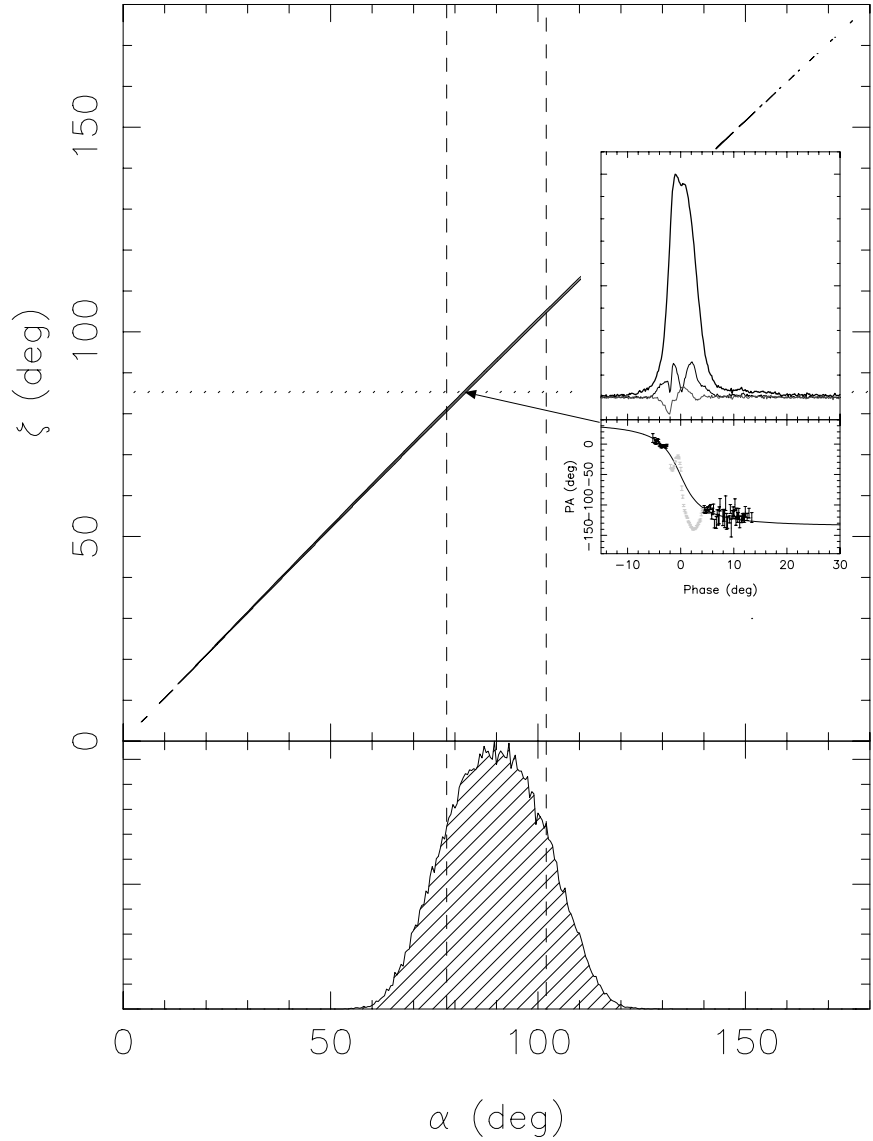}
\caption{\label{fig:profile} System geometry for PSR J2222$-$0137 as derived from
a least-squares-fit of the Rotating Vector Model (RVM) to the 
PA of the linearly polarized emission. The top panel shows the 
regions of the magnetic
inclination angle ($\alpha$) and viewing angle ($\zeta$) plane
with the best RVM fits (see inset and text for details).
We also mark the constraint on the orbital inclination angle as an
horizontal strip. Also, assuming a filled emission beam,
we derive a distribution of inclination angles (lower panel) that is
consistent with the observed pulse width (see text for details).
The vertical dashed lines indicate a $\pm 12^\circ$ band around $90^\circ$.
For the point that satisfies the polarimetric and orbital
inclinations constraints ($\alpha \, = \, 83^\circ$ and
$\zeta\, = \, 85.27^\circ$) we calculate the RVM PA versus phase curve and
superpose it on the measurements (see bottom plot of inset).}
\end{figure*}

In Figure~\ref{fig:profile}, we display (in the inset, top panel) the
pulse profile for PSR~J2222$-$0137. This was obtained from the addition of the
best Effelsberg detections of the pulsar, which were cleaned of RFI and
then calibrated using the noise diode observations made every hour
during the observations. This was done using the
``PAC'' program, which is part of {\tt PSRCHIVE}
using the ``Single Axis'' model.
We  corrected for the Faraday rotation of the pulsar using the
rotation measure published in Paper I
($\rm RM \, = \, 2.6(1)\, rad \, m^{-2}$).

In the bottom panel of the inset, we display
the PA of the linear polarization as a function of spin phase.
This is measured in a counter-clockwise
sense, starting from North through East. Furthermore, a
system with an orbital inclination of zero degrees has its angular momentum
pointing towards the Earth. This is the ``observer's convention''
defined in \cite{tempo22} that is assumed in {\tt PSRCHIVE}.

To this observed PA curve, we have tried to fit a rotating vector model
(RVM, \citealt{RVM}). This is not a clean fit: the PA profile
is clearly more complex than described by a simple RVM model, so the
interpretation is not straightforward. This is a situation
that has been long observed among MSPs (e.g., \citealt{Xilouris1998}).

In order to obtain a meaningful model, we chose two longitude
ranges (highlighted in black, ignored regions in gray), which avoid a pulse region that can be associated with so-called "core" emission, which is known to commonly disturb a smooth PA swing (\citealt{Rankin1983}).
This allows RVM fits to determine a viewing angle ($\zeta$,
the angle between the spin axis and the magnetic axis) that
is similar to the orbital inclination of the system, $i$. This is
expected given the fact that this pulsar is recycled: it was spun up
with gas from its companion, this implies that there was a transfer
of orbital angular momentum to the pulsar. This implies that after
accretion the pulsar's spin angular momentum was parallel to the
orbital angular momentum. We see no signs of a disruptive event (like a second SN)
that might have changed the orbital plane after the recycling episode,
so the spin angular momentum should still be closely aligned with
the orbital angular momentum.

In the main panel, we display, for each value of
magnetic inclination angle ($\alpha$, the angle between the spin axis and
the magnetic axis) and viewing angle $\zeta$ 
the quality of the RVM fit to the PA curve. At each point we keep $\alpha$ and $\zeta$ fixed, while minimizing the $\chi^2$-value in a least-squares fit to the reference phase $\phi_0$ and reference PA
$\psi_0$.  The inclined contour lines indicate the $1-\sigma$ region. Clearly, there is
a strong correlation between $\alpha$ and $\zeta$, which is a well known effect for a small pulse duty cycle as observed here (e.g.~\citealt{handbook}).

In order to constrain the geometry further, we
mark the constraint on the orbital inclination angle
($\zeta \, = \, i\, = \, 85.27^\circ$), which is derived from the Shapiro delay
and the measurement of $\dot{x}$ (see section~\ref{sec:results}),
as an horizontal strip. This is the aforementioned constraint that
the spin angular momentum of the pulsar is parallel to the orbital angular momentum. 

Assuming a filled emission beam, we derive a distribution of magnetic
inclination angles (lower panel) that is consistent with the observed pulse width.
The vertical dashed lines indicate a $\pm 12^\circ$ region around
$90^\circ$. As we can see, the region given by the intersection of
the area allowed by the polarimetry and the measurement of the
orbital inclination is well within the allowed region of
magnetic inclination angles; this self-consistency suggests that the
choice of longitude ranges we have made and the best-resulting RVM fit
(Also displayed in the bottom panel of the inset, superposed to the
measurements of the linear PA) is not entirely arbitrary, providing
a reasonable estimate of $\alpha\, \sim \, 83^\circ$. This means that
the magnetic pole passes only within a couple of degrees of the 
line of sight.

\subsection{Data reduction for pulsar timing}

All dedispersed pulse profiles were added in blocks lasting 10 minutes and for sub-bands
of 128 MHz for NRT NUPPI, 150/125 MHz for the Effelsberg PSRIX data and 80 MHz for
the Lovell ROACH data (see Table~\ref{table:observations}).
They were then cross-correlated with a low-noise template using
the Fourier routine described in \cite{TaylorRG} and implemented in the ``PAT''
routine of {\tt PSRCHIVE} to derive the
topocentric pulse times of arrival (TOAs).
We then used {\tt tempo}\footnote{\url{http://tempo.sourceforge.net/}}
to correct the TOAs using the telescopes' clock
corrections and to convert them to the Solar System barycenter. This program
reports all time-like units in Dynamical Barycentric time. To do this, the
motion of the radio telescopes relative to the Earth was calculated using the
data from the International Earth Rotation Service, and to the barycenter using
the DE430 solar system ephemeris \citep{Folkner2014}.
Finally, the differences between the measured TOAs and the predictions of
our timing model (the {\em residuals}) were minimized using {\tt tempo} 
by varying the parameters in the model, with and without a position fit.
The parameters that best fit the data are presented in
Table~\ref{table:timing}. To model the binary orbit, we used
the DD model described by \cite{Damour85,Damour86}. These use the
``range'' ($r$) and ``shape'' ($s$) parameters to quantify the Shapiro delay;
for high inclinations their correlation as small as those of the
orthometric amplitude $h_3$ and ratio $\varsigma$ in the orthometric
parameterization of the Shapiro delay \citep{FreireWex}.

\begin{figure*}
\centering
\includegraphics[width=0.8\textwidth]{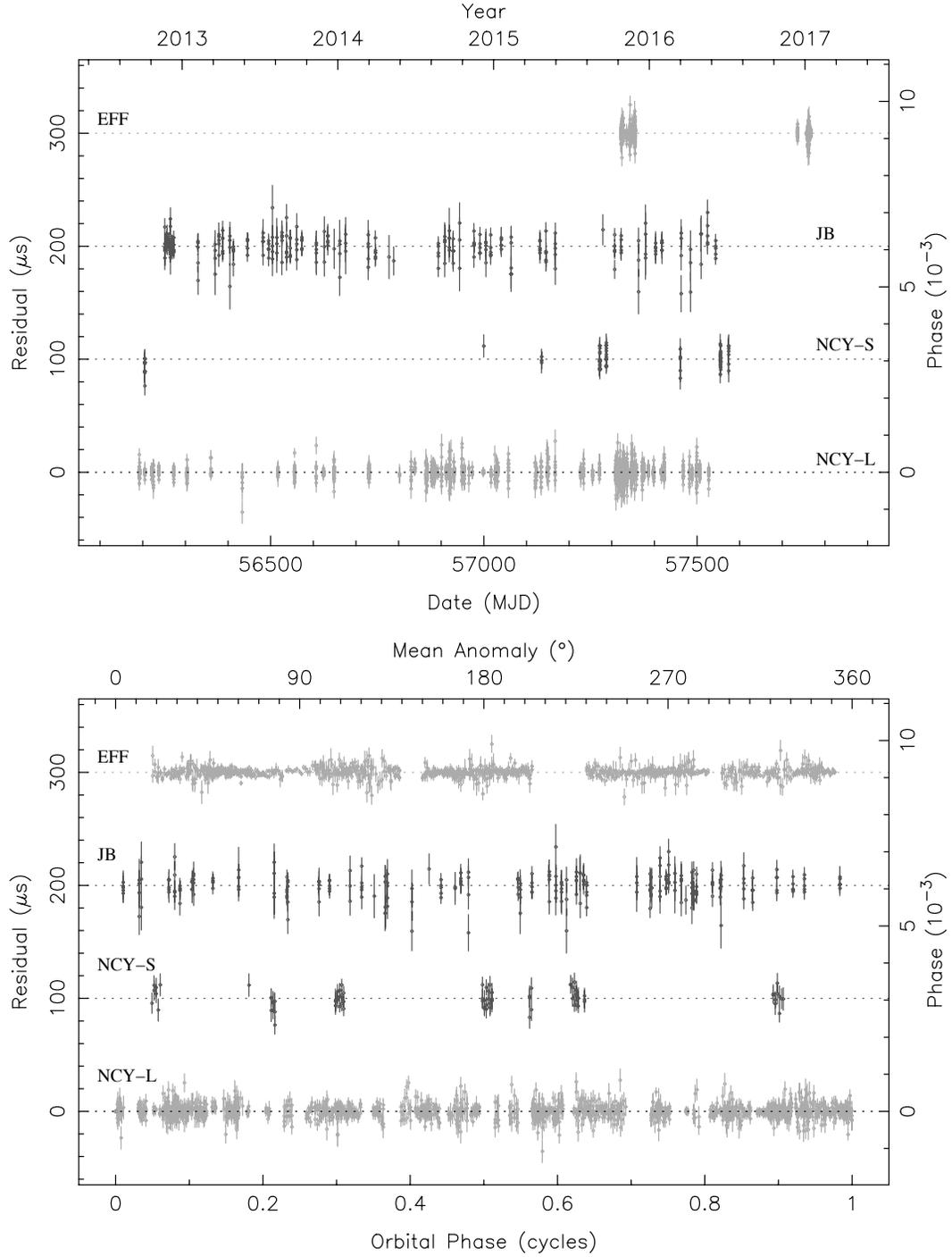}
\caption{\label{fig:residuals} Timing residuals (measured 
pulse arrival times $-$
model pulse arrival times) as a function of MJD (top) and mean anomaly (bottom)
for the timing solution of PSR J2222$-$0137 where we fit for position.
The residuals are displayed with different offsets for each
instrument. No trends are noticeable in the residuals, indicating that
the model describes the topocentric pulse times of arrival
(TOAs) well. EFF - Effelsberg TOAs, JB - Lovell telescope (Jodrell Bank),
NCY-S - Nan\c{c}ay at S-band and NCY-L - Nan\c{c}ay at L-band.}
\end{figure*}

The residuals associated with the best-fit model are displayed in 
Figure~\ref{fig:residuals}. This timing model fits for right ascension
($\alpha$) and declination ($\delta$), with no trends
detectable in the residuals. The residual root mean square 
(rms) for the overall solution is $3.48\,\mu$s, this represents a
fraction of $\sim 10^{-4}$ of the spin period.
For the individual observing systems the residual rms are presented
in Table~\ref{table:observations}, where we also listed the 
times added in quadrature (EQUAD) to each data set in order to
achieve a reduced $\chi^2$ of 1.  This procedure results in more
conservative uncertainty estimates for all timing parameters.

\section{Astrometry}
\label{sec:astrometry}

We now discuss the astrometric parameters derived from timing.
First, we fitted for parallax, proper motion and position simultaneously.
In this way, we were able to make an independent measurement of these
parameters using only our data set, i.e., not taking into account the early VLBI
position (since there might be systematic offsets between the positions at
the reference epoch). For the parallax we obtained a value of 
$3.88\, \pm\, 0.31$ milliarcseconds (mas), which is consistent with VLBI 
parallax ($3.742^{+0.013}_{-0.016}\, \rm mas$). Since the latter is 
$\sim 20$ times more precise, we kept it fixed in all subsequent work.

We then fitted for proper motion, still fitting for
position at the same time. Doing this, we obtained
$\mu_{\alpha} \, = \, 45.00 \, \pm \, 0.21 \rm \, mas\, yr^{-1}$ and
$\mu_{\delta} \, = \, -6.35 \, \pm \, 0.47 \rm \, mas\, yr^{-1}$.
Both values are $\sim$1-$\sigma$ consistent with the proper motions from Paper~II,
however the latter are still more precise by an order of magnitude;
for this reason we used the VLBI proper motions in all subsequent fits.

Finally, we kept the parallax and proper motion fixed at the VLBI values
and fitted for right ascension ($\alpha$) and declination ($\delta$), we obtained
$\alpha \, = \, 22^{\rm h}\, 22^{\rm m}\, 05\fs969080(13)$ and
$\delta \, = \, -01^{\circ}\, 37' \, 15\farcs7262(5)$ for MJD = 55743;
these represent offsets of $d \alpha \, = \, 0.000021(13)$
seconds and $d \delta\, = \, -1.8\,\pm\,0.5\,$mas
relative the VLBI position. This means the timing position in $\alpha$ is 2-$\sigma$
consistent with the $\alpha$ from VLBI, but in $\delta$ the offset is nearly $4$-$\sigma$
significant: fitting for the position produces a large reduction in the $\chi^2$ of the fit
(from 3076.9 to 2874.0). This reduction is too large to ignore, so we investigated
the reasons for this.

\begin{enumerate}
\item We use the same epoch for the position as the epoch quoted in
Paper II, thus verifying that the proper motion is not the cause of the positional offset.
\item This offset is not due to any misalignment of the reference frames: According to
\cite{Folkner2014}, the DE430 reference frame is aligned to the
International Celestial Reference frame to better than 0.2 mas
(10 times smaller than the measured offset).  Nor is it caused by shortcoming of
{\tt tempo}: A fit with {\tt tempo2} \citep{tempo2,tempo22} results in a very similar offset.
\item Since the pulsar is within 8 degrees of the Ecliptic, we have investigated
whether the solar wind might have a significant effect on the timing.
The position of the pulsar can change significantly
with the assumed solar wind density parameter, but in no case does it match
the VLBI position. Furthermore, density parameters for the solar wind
higher than 10 cm$^{-3}$ cause a significant degradation
in the timing model.
\item Nevertheless, it is known that the solar wind density changes with time,
implying that, to time a pulsar near the ecliptic properly, one would in principle
need to know how the solar wind density changes with time. This information
is not readily available. To avoid this problem, we adopted the common practice in
pulsar timing arrays (PTAs) and excluded all TOAs where the Sun is within $15^\circ$
of the pulsar, i.e., where the Ecliptic longitude of the Sun
$\lambda_{\odot}$ is within $12.7444^\circ$ of the longitude of the pulsar,
$\lambda_{\rm PSR}\, = \, 336.7319^\circ$.
This corresponds to any days from Feb. 13 to March 11. Doing this
does not change the positional offset.
\item Independently of this, many pulsars show variations in their DMs
caused by their motions relative to the Earth. The line of sight to
the pulsar goes through varying electron column densities, causing
an irregular (and unpredictable) change in their DMs. These changes
affect especially timing parameters with a long signature, like
the astrometric parameters.\\
If we model DM variations using a piecewise-constant function
(the so-called ``DMX'' model, \citealt{Demorest2013})
and keep the VLBI positions, we can see an apparently
yearly change in the DM.
However, if we used the DMX model {\em and} fit for position, then we see that
the best fit is where the position changes and the DMs change approximately
linearly with time, with no sign of a yearly modulation.
This means that when we used the DMX model and kept the VLBI position, the DM
coefficients were in effect absorbing the yearly residuals.
\end{enumerate}

Given this situation, we conclude that a) we get a perfect agreement
with the parallax, and a reasonable agreement with the
proper motion in Paper II, b) the positional offset we
found is qualitatively and quantitatively robust, and that
c) it is not an effect of DM variations, caused by the Sun or otherwise.
The latter point is corroborated by the fact that our measurement
of parallax coincides with the parallax measurement in Paper II;
the parallax is the parameter that suffers most from timing systematics
with a yearly signature.

Could some systematic effects have affected the VLBI parallax?
According to A. Deller (2017, private communication), there are two potential sources
of error that were neglected in Paper II by the assumption that the out-of-beam
calibrator position uncertainty dominates the target absolute position uncertainty:
\begin{enumerate}
\item Core-shift in the out-of-beam calibrator J2218$-$0335 (since its position is
defined at 2.3/8.4 GHz, and the PSR J2222$-$0137 measurements were done at 1.6 GHz).
This error cannot be evaluated
precisely - that would require a careful, multi-frequency/multi-calibrator campaign - but potentially it
could provide a shift of the order of 1 mas \citep{Sokolovski2011}.
\item Errors in the residual phase-referencing errors from the out-of-beam calibrator to the in-beam calibrator.
The angular separation here is $\sim\, 2^\circ$, which could originate an error of $\sim$1-2 mas in a typical 
observation, but could be even more if ionospheric conditions are unfavorable. This effect should
average out, but a fraction of a mas is to be expected for the final position. Again, verifying this
would require a careful, multi-frequency/multi-calibrator campaign.
\end{enumerate}
These contributions are systematically taken into account in recent publications. As an example,
\cite{Deller2016} estimated an  absolute uncertainty of the in-beam calibrator
for PSR~J2145$-$0759 of about 2.5 mas. This should be smaller for PSR~J2222$-$0137
because the distance between pulsar and calibrator is 30\% smaller and the latter pulsar
is not so far South,  but nevertheless it is conceivable that these effects could produce
an offset similar to what we observe ($\sim \, 2 \, \rm mas$), although without a dedicated multi-frequency/multi-calibrator campaign
it will not be possible to quantify this more precisely. Nevertheless, these recent
uncertainty estimates show that the 0.1 mas precision for the absolute position
claimed in Paper II is too optimistic.

In any case, the two timing solutions presented in Table~\ref{table:timing} -- one
using the VLBI position and the other using our best fit position -- show that
the Shapiro delay parameters and the orbital variability parameters change by 1 $\sigma$
or less with these different positions, i.e., the masses derived in this paper are consistent
for these two positions.

\begin{table*}
  \caption{Parameters for the PSR J2222$-$0137 binary pulsar}
  {\scriptsize
  \begin{center}
  \renewcommand{\arraystretch}{1.0}
  \begin{tabular}{lcc}
  \hline
  \multicolumn{3}{l}{\bf General timing parameters}\\
  %\hline
  Right Ascension, $\alpha$ (J2000) \dotfill & 22:22:05.969101 (a)  & 22:22:05.969080(13) \\
  Declination, $\delta$ (J2000) \dotfill & $-$01:37:15.72441 (a) & $-$01:37:15.7262(5) \\
  Proper motion in $\alpha$, $\mu_{\alpha}$ (mas\,yr$^{-1}$) \dotfill & 44.73 (a) & 44.73 (a)  \\
  Proper motion in $\delta$, $\mu_{\delta}$ (mas\,yr$^{-1}$) \dotfill & $-$5.68 (a) &  $-$5.68 (a) \\
  Parallax, $\pi_x$ (mas)  \dotfill & 3.742 (a) & 3.742 (a)  \\
  Spin frequency, $\nu$ (Hz) \dotfill & 30.47121380904560(27) & 30.47121380904550(29) \\
  Spin frequency derivative, $\dot{\nu}$ ($10^{-18}$ Hz\,s$^{-1}$) \dotfill & $-$5.38780(28) & $-$5.38746(34) \\
  Dispersion Measure, DM (\dmu) \dotfill & 3.277 & 3.277 \\
  DM derivative , DM1 (\dmu\,$\rm yr^{-1}$) \dotfill & 0.00082 & 0.00082 \\
  FD1 \dotfill & $-$0.00135(14) &  $-$0.00131(13)  \\
  FD2 \dotfill &    0.0048(4)   &     0.0046(4) \\
  FD3 \dotfill & $-$0.0069(6)   &  $-$0.0068(6) \\
  FD4 \dotfill &    0.00331(28) &     0.00323(26) \\
  Rotation Measure, RM (rad m$^{-2}$) \dotfill & 2.6(1) (b) & 2.6(1) (b) \\
  Weighted residual rms ($\mu$s) \dotfill & 3.481 & 3.365 \\
  $\chi^2$ \dotfill & 3076.9 & 2874.0 \\
  Reduced $\chi^2$ \dotfill & 1.076 & 1.006\\
   \multicolumn{3}{l}{\bf Binary Parameters}\\
  Orbital Period, $P_b$ (days) \dotfill & 2.44576456(13) & 2.44576469(13) \\
  Projected Semi-major Axis of the pulsar orbit, $x$ (lt-s) \dotfill & 10.8480229(6) & 10.8480239(6) \\
  Epoch of Periastron, $T_0$ (MJD) \dotfill & 56001.38392(7) & 56001.38381(8) \\
  Orbital Eccentricity, $e$ \dotfill & 0.000380940(3) & 0.000380967(30) \\
  Longitude of Periastron, $\omega$ ($^\circ$) \dotfill & 119.916(11) & 119.900(11) \\
  Rate of advance of Periastron, $\dot{\omega}$  \dotfill & 0.1004(28) & 0.1033(29) \\
  Shapiro delay "shape" $s$  \dotfill &  0.99669(29) & 0.99659(30) \\
  Shapiro delay "range" $r$ (\Msun) \dotfill & 1.323(25) & 1.293(25) \\
  Variation of $P_b$, $\dot{P}_{b,\rm obs}$ ($10^{-12}\,\rm s\,s^{-1}$) \dotfill & 0.27(9) & 0.20(9) \\
  Variation of $x$, $\dot{x}$ ($10^{-15} \rm lt$-$\rm s\,s^{-1}$) \dotfill & 7.8(30) & 3.5(30) \\
  Position angle (PA) of line of nodes, $\Omega$ ($^\circ$, J2000) \dotfill & \multicolumn{2}{c}{$5^{+15}_{-20}$ (a)} \\
  %\hline
  \multicolumn{3}{l}{\bf Derived Parameters}\\
  %\hline
  Galactic Longitude, $l$ ($^\circ$) \dotfill & \multicolumn{2}{c}{62.0184} \\
  Galactic Latitude, $b$ ($^\circ$) \dotfill &  \multicolumn{2}{c}{$-$46.0753} \\
  Ecliptic Longitude, $\lambda$ ($^\circ$) \dotfill &  \multicolumn{2}{c}{336.7319} \\
  Ecliptic Latitude, $\beta$ ($^\circ$) \dotfill &  \multicolumn{2}{c}{7.9771} \\
  Distance, $d$ (pc) \dotfill & \multicolumn{2}{c}{267.3 (a)} \\
  Total proper motion, $\mu_{T}$ (mas\,yr$^{-1}$) \dotfill & \multicolumn{2}{c}{45.09(2) (a)} \\
  PA of proper motion, $\Theta_{\mu}$ ($^\circ$, J2000) \dotfill & \multicolumn{2}{c}{97.23 (a)} \\
  Transverse velocity, $v_{T}$ (km\,s$^{-1}$) \dotfill & \multicolumn{2}{c}{$57.1^{+0.3}_{-0.2}$ (a)} \\
  Pulsar Spin Period, $P$ ($\s$) \dotfill & 0.03281785905434272(30) & 0.03281785905434283(31) \\
  Spin Period Derivative, $\dot{P}$ ($10^{-21}$ s\,$\ps$) \dotfill & 58.0272(30) & 58.0236(36) \\
  Intrinsic Period Derivative, $\dot{P}_{\rm int}$ ($10^{-21}$ s\,$\ps$) \dotfill &  \multicolumn{2}{c}{17.50} \\
  Surface Magnetic Field Strength, $B_0$ ($10^{9}$ G) \dotfill & \multicolumn{2}{c}{0.76} \\
  Characteristic Age, $\tau_c$ (Gyr) \dotfill & \multicolumn{2}{c}{29.7} \\
  Spin-down energy, $\dot{E}$ ($10^{30}$ erg s$^{-1}$) \dotfill & \multicolumn{2}{c}{19.5} \\
  Mass function, $f$ (\Msun) \dotfill &  0.22914228(4) & 0.22914232(4) \\
  Pulsar mass, $m_{p}$ (\Msun) \dotfill & 1.84(6) & 1.76(6) \\
  Total binary mass, $M$ (\Msun) \dotfill & 3.16(8) & 3.05(9) \\
  Orbital inclination, $i$ ($^\circ$)  \dotfill & 85.34(21) & 85.27(22) \\
  Intrinsic $\dot{P}_b$, $\dot{P}_{b,\rm int}$ ($10^{-12}\,\rm s\,s^{-1}$) \dotfill & +0.01(9) & $-$0.06(9) \\
  \hline
 \end{tabular}
 \end{center}
 \tablecomments{Timing parameters derived using {\tt tempo}. The left column is the timing solution
 derived using the position from Paper II. The right column is the timing solution
 where we fit for position ($\alpha$ and $\delta$). The reference
 epoch is MJD = 56000, the position epoch is for MJD = 55743, as in Paper II.
 We use the DE 430 Solar System ephemeris, time units are in barycentric dynamic time
 (TDB). The FDN parameters model non-dispersive changes in the average residual with
 frequency \citep{Arzoumanian15}. The binary parameters derived from the timing (all but $\Omega$)
 are relative to the DD orbital model. Numbers in parentheses
 represent 1-$\sigma$ uncertainties in the
last digits as determined by the timing programs, scaled such that the reduced
$\chi^2= 1$. (a) Value determined from VLBI measurements. (b) Value from Paper I. \label{table:timing}} }
\end{table*}

\section{Timing Results}
\label{sec:results}

In Table~\ref{table:timing}, we present our main results, the timing parameters
for PSR~J2222$-$0137.  As mentioned in Section~\ref{sec:astrometry}, we have
discarded all TOAs taken when the Sun is at an angular distance
from the pulsar smaller than 15 degrees. 
Furthermore, as mentioned in the previous section, when we fit for position
and use the DMX model, we see a steady secular increase in the DM, for this
reason we decided to model the DM evolution with a DM and DM derivative only.
In a later stage we turn off the fit for both DM and its derivative since
we also fit for the FD paramaters  \citep{Arzoumanian15} - If all of these parameters
are fit together, the DM can easily change to unrealistically large or small
(sometimes negative) values. The uncertainties presented
are 68\% (1-$\sigma$ equivalent) confidence limits derived by {\tt tempo}.
We now discuss the significance of some of these parameters.

%------------------------------------------------------------------------------------------

\subsection{Spin parameters}

The spin properties for this pulsar were discussed in Paper I.
Most of the observed spin period
derivative, ($\dot{P}\, = \, 5.80\, \times 10^{-20}$) results from
kinematic contributions. Once those are subtracted, the resulting intrinsic spin-down
is $\dot{P}_{\rm int} \, = \, 1.75 \, \times 10^{-20}\, \rm s \, s^{-1}$. This is slightly
larger than the value presented in Paper I because,
apart from correcting for the Shklovskii effect \citep{Shklovskii70}, we also
correct for the difference in Galactic accelerations of PSR~J2222$-$0137
and the Solar System \citep{DamourTaylor91} using the latest
model for the rotation of the Galaxy \citep{reid2014}. However, the latter is a
small term that does not change any of the basic conclusions: a
characteristic age $\tau_c\, =\, P / (2 \dot{P}_{\rm int})\, \sim\, 30$ Gyr
and a relatively small B-field ($B_0\,= \, 3.19 \, \times\,  10^{19}\, {\rm G}\,\sqrt{P
\dot{P}_{\rm int}} \, = \, 7.6 \, \times \, 10^8$G;
see e.g., \citealt{handbook}).

%------------------------------------------------------------------------------------------

\subsection{Shapiro delay}

The masses of the components of the binary can be determined reasonably well
using the Shapiro delay alone: we obtain $m_p\,=\, 1.76(6)\, \Msun$, $m_c\, = \,
1.293(25)\, \Msun$ and $i\,= \,85.27(22)^\circ$ or $i'\,= \,94.73(22)^\circ$.
If we don't fit for the position, the masses are 1 $\sigma$ larger
(see Table ~\ref{table:timing}).

\begin{figure*}
\centering
\includegraphics[width=\textwidth]{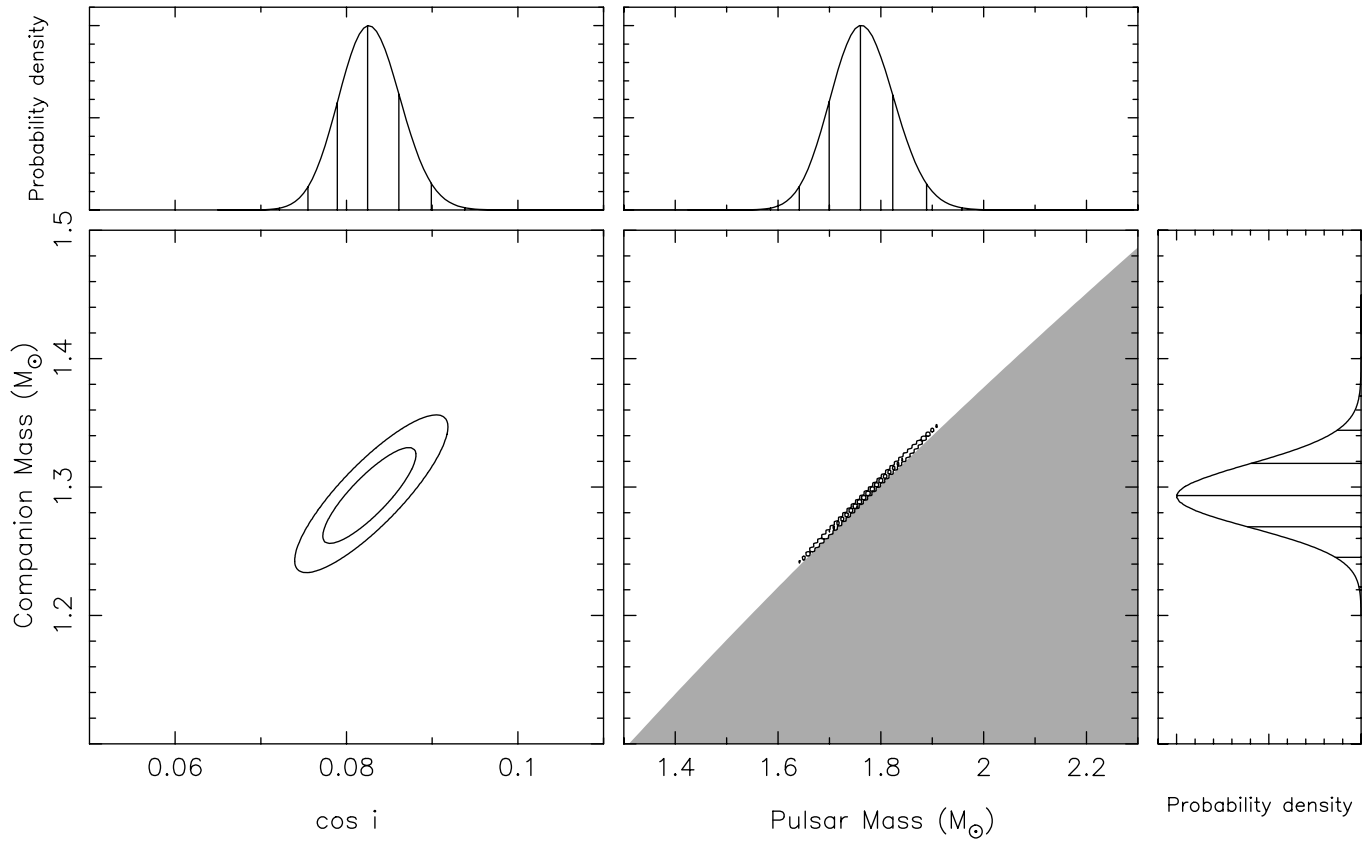}
\caption{\label{fig:mass_mass} Current constraints from timing of PSR
J2222$-$0137. The black, solid contour lines include 68.3 and 95.4\% of the 2-D
probability distribution function derived from our $\chi^2$ maps (see text).
{\em Left}: $m_c$ - $\cos i$ plane. {\em Right}: $m_c$-$m_p$ plane.
The gray region is excluded by the mathematical constraint $\sin i \leq 1$.
Top and side panels: 1-D probability distribution functions for 
$\cos i$, $m_p$ and $m_c$, with median, 1, 2 and 3 $\sigma$ equivalent
percentiles indicated as vertical lines. The upper mass limit for a rigidly
rotating WD is $1.48\,\Msun$.}
\end{figure*}

In order to verify these values and their uncertainties
we have made a $\chi^2$ map of the $\cos i - m_c$ space. At each point, we fix the
Shapiro delay values and fit for all other parameters, recording
the $\chi^2$ for each fit. From this $\chi^2$ map we derive
a 2-D probability density map (see Fig.~\ref{fig:mass_mass})
according to the Bayesian procedure described in \cite{Splaver+02},
see e.g., \cite{Barr2017} for details.
From this, we obtain $m_p \, = \, 1.760^{+0.063}_{-0.061} \, \Msun$,
$m_c \, = \, 1.293^{+0.025}_{-0.024} \, \Msun$ and
$i\, = \, 85.27(21)^\circ$ (or $i' \, = \, 94.73(21)^\circ$).
A bootstrap Monte-Carlo estimate yields smaller uncertainties, e.g.,
$m_c \, = \, 1.293(17) \, \Msun$, for this reason we keep the {\tt tempo}
estimate since it is more conservative.
These values are not consistent with the masses published in Paper~I, in terms
of the latter's estimated values and 1-$\sigma$ uncertainties, they are both
4$\,\sigma$ too large. It is important to understand the reason for this
difference.

We begin by pointing out that we
have made a substantially larger number of measurements with much better
orbital coverage over a more extended period of time, all of which
are important not only for improving the precision of the Shapiro delay,
but also for reducing the correlations between different parameters.
One of these correlations is especially important, and had already been
explicitly mentioned in Paper~I: it is the correlation between the Shapiro delay
parameters and the rate of advance of periastron, $\dot{\omega}$.
Indeed, in Paper~I, when the authors fit for $\dot{\omega}$,
they obtain a significant improvement in the quality of their fit
(an early indication that the effect is measurable) and an increase in
the masses derived from the Shapiro delay by about 1 $\sigma$ compared
to their tabulated value (where this effect was not taken into account).
This leads us to conclude that our measurement of
$\dot{\omega}$ is one of the main reasons for the large difference
in the Shapiro delay measurements.

%------------------------------------------------------------------------------------------

\subsection{Advance of periastron}
\label{sec:omega-dot}

In this timing campaign, we have measured $\dot{\omega}$ with high
($35 \, \sigma$) significance: $\dot{\omega}_{\rm obs} = 0.1033(29)^\circ \rm
yr^{-1}$.
Given the optical non-detection of the companion to PSR~J2222$-$0137,
it is clear that it must be a highly compact object. Like the pulsar,
it will behave essentially like a point mass. This means that ``classical"
rotational or tidal contributions to the observed $\dot{\omega}$
should be absent. Therefore, $\dot{\omega}_{\rm obs}$ should
be dominated by the relativistic contribution, with an insignificant
contribution from kinematic effects (section~\ref{sec:kinematic}).

For all known systems where the masses can be determined independenty, 
general relativity (GR) provides an accurate prediction of this quantity;
to leading post-Newtonian order it depends only on the Keplerian
parameters and the total mass of the system $M$, in solar masses
(\citealt{Robertson1938,Weisberg1982}):
\begin{equation} \label{eq:omdot}
  \dot{\omega}_{\rm GR} = 3 \frac{(M{\rm T}_{\odot})^{2/3}}{1- e^2} 
  \left( \frac{P_b}{2\pi } \right)^{-5/3},
\end{equation}
where  ${\rm T}_{\odot} \equiv ({\cal GM})_\odot^{\rm N} c^{-3} = 4.9254909476412675(...)\, \rm \mu
s$ is the nominal solar mass parameter\footnote{
Although neither $G$ nor the mass of the Sun ($\Msun$) are known to better than 4 
decimal places, their product is known to more than nine decimal places. Recently, the IAU 
2015 Resolution B3 has defined the {\em nominal solar mass 
parameter}, denoted by $({\cal GM})_\odot^{\rm N}$, to be {\em exactly}
$1.3271244 \times 10^{20} \rm \, m^3\, s^{-2}$. Thus the value
for $\rm T_{\odot}$ as defined above is also exact, 
it is no longer tied to the actual (time-varying) mass of the Sun, but
instead to the SI units of time and length.}
in time units, $c$ is the speed of light.
Using the $M$ presented in Paper~I, we
should expect $\dot{\omega}_{\rm GR, 1} = 0.077^\circ \rm yr^{-1}$; for the
$M$ presented in this paper ($3.05(9)\, \Msun$) we should expect
$\dot{\omega}_{\rm GR, 2} = 0.0943^\circ \rm yr^{-1}$.
The observed value is about 3 $\sigma$ above $\dot{\omega}_{\rm GR, 2}$,
while it is 9 $\sigma$ above $\dot{\omega}_{\rm GR, 1}$. This is an
independent indication that the larger masses derived in this paper are closer to
the real value, however the agreement with  $\dot{\omega}_{\rm GR, 2}$
is not very good either. We should keep in mind that for low-eccentricity
orbits this quantity is easily affected by systematic effects.
As an example of this, if we use the DMX model and fit for position
then we obtain $\dot{\omega}_{\rm obs} = 0.1006(35)^\circ \rm
yr^{-1}$, if we don't fit for position then
$\dot{\omega}_{\rm obs} = 0.1001(35)^\circ \rm yr^{-1}$, these values
are 1 $\sigma$ smaller; they are only 1.7 $\sigma$ above $\dot{\omega}_{\rm GR, 2}$.
As a comparison, the Shapiro delay
masses vary by only $\pm \, 0.1 \,\sigma$ respectively.

We can in principle combine $\dot{\omega}_{\rm obs}$ with the
Shapiro delay to obtain self-consistent GR mass
measurements, using the DDGR model \citep{Damour86}.
However, because we believe that the $\dot{\omega}_{\rm obs}$
is currently contaminated by systematic effects, we don't
regard the result of such a fit as being reliable.

However, this situation will change. As we extend our timing baseline $T$ the
uncertainty of $\dot{\omega}_{\rm obs}$ will 
decrease at a rate given by $T^{-3/2}$; we also expect that the
relative contribution from systematic effects will decrease. This
means that a self-consistent combination of $\dot{\omega}_{\rm obs}$ with
the Shapiro delay will provide much more precise masses in the future,
as for PSRs~J1903+0327, J1807$-$2500B,
J0453+1559 and J1946+3417 \citep{FreireJ1903,J1807,Martinez2015,Barr2017}. 

%------------------------------------------------------------------------------------------

\subsection{Kinematic effects}
\label{sec:kinematic}

Given the well-known values for distance and proper motion, we can estimate the
kinematic contributions to the observed variations of $P_b$
\citep{Shklovskii70,DamourTaylor91}, $x$ and $\omega$ \citep{Arzoumanian96,Kopeikin}.

\subsubsection{Variation of the orbital period}
\label{sec:Pbdot}

For $\dot{P}_b$, we expect a kinematic contribution of $\dot{P}_{b,k} \, = \,
0.2648(17) \times 10^{-12} \, {\rm s\,s^{-1}}$, which is due to the
Shklovskii effect ($\dot{P}_{b,Shk} \, = \,0.2790(11) \times 10^{-12} \, {\rm s\,s^{-1}}$
- the uncertainty here is dominated by the uncertainty in the distance to the system)
and the Galactic acceleration
($\dot{P}_{b,Gal} \, = \,-0.0142(13) \times 10^{-12} \, {\rm s\,s^{-1}}$, where we
assumed a 10\% uncertainty in the vertical acceleration of the system).
The GR prediction for the orbital decay caused by the emission of gravitational waves,
$\dot{P}_{b,{\rm GR}} 
\, = \, -0.0077(4) \times 10^{-12}\,{\rm s\,s^{-1}}$, is 34 times smaller than
the kinematic contribution. The total predicted $\dot{P}_b$ is then 
$\dot{P}_{b,p} \, = \, 0.2571(17) \times 10^{-12}\,{\rm s\,s^{-1}}$.

From our data we get $\dot{P}_{b, \rm obs} = (0.20 \pm 0.09) \times
10^{-12}  \rm s \, s^{-1}$, which is consistent with $\dot{P}_{b,p}$.
Subtracting the expected kinematic contribution, we  obtain the intrinsic 
component of the orbital variation, 
$\dot{P}_{b, \rm int} \, = \, -0.063 \, \pm \, 0.085 \times 10^{-12}\,{\rm s\,s^{-1}}$.
This is consistent, but not yet precise enough for a detection of the quadrupolar GW
emission predicted by GR.

Subtracting the expected GR contribution, we obtain
an excess of $\dot{P}_{b, \rm xs} \, = \, -0.055 \, \pm \, 0.085 \times 10^{-12}\,{\rm s\,s^{-1}}$,
which represents the upper limit for dipolar GW emission.
As discussed in section~\ref{sec:GWs}, this
is already small enough to introduce interesting constraints on the emission of
dipolar gravitational waves from this system. Furthermore, these limits will
improve fast since the uncertainty of $\dot{P}_b$ decreases with $T^{-5/2}$.

\subsubsection{Secular variation of the projected semi-major axis of the pulsar's orbit}

The expected kinematic effect on the projected semi-major axis of the pulsar's
orbit ($x$) is given by \citep{Kopeikin}:
\begin{equation}
\frac{\dot{x}}{x} \, = \, \mu_{T} \cot i \sin \left( {\Theta_{\mu}} - \Omega \right),
\label{eq:xdot}
\end{equation}
where $\Theta_{\mu}$ is the PA of the proper motion ($97.23^\circ$)
and $\Omega$ is the PA of the line of nodes; for these
angles we use the ``observer's convention'' mentioned in
Section~\ref{sec:polarimetry}\footnote{In \cite{FreireJ1903},
eq.~\ref{eq:xdot} has a minus sign, which corresponds to a
convention where PAs are measured from North through East
and a binary system with an orbital inclination of $0^\circ$ has
its angular momentum pointing {\em away} from us. Such a
convention is left handed, the convention used here is right
handed.}.
Despite the relatively large $\mu_{T}$, the expected $\dot{x}$ is
relatively small mostly because $i$ is close to
$90^\circ$; this implies that $\cot i$ term is also small.
Nevertheless, our measurement of $\dot{x}$ is already
precise enough to start introducing some constraints.
As mentioned in Section~\ref{sec:introduction}, $\Omega$ was
measured in Paper II and is $5_{-20}^{+15}$$^\circ$, this means that we
should expect $\dot{x}_p \, = \, \pm 6.06\, \times \, 10^{-15}$ lt-s s$^{-1}$,
depending on whether $i\,= \,85.27(22)^\circ$ or $i'\,= \,94.73(22)^\circ$
respectively. The observed value,
$\dot{x}_{\rm obs} \, = \, +(3.5\, \pm \, 3.0) \, \times \, 10^{-15}$ lt-s s$^{-1}$,
is 1-$\sigma$ consistent with the $\dot{x}_p$ for 
$i$ and 3.2 $\sigma$ larger than the $\dot{x}_p$ for $i'$.
From this we conclude
that $i$ is the more likely the orbital inclination.

The uncertainty in $\dot{x}_{\rm obs}$ decreases with $T^{-3/2}$;
this implies that soon it will provide a choice of the
orbital inclination with much higher confidence.

\subsubsection{Annual orbital parallax}

Apart from the secular variation of $\dot{x}$ there is
another kinematic effect on $x$, the annual orbital parallax.
This is a yearly modulation of $x$ caused by the
the changing viewing angle of the pulsar's orbit
due to the Earth's orbital motion \citep{Kopeikin}.

The DDK orbital model in {\tt tempo} can fit for this
and other annual effects, like the orbital parallax,
but it automatically takes $\dot{x}$ into account.
It is therefore not easy to separate the constraints
introduced by the annual orbital parallax from
the constraints introduced by $\dot{x}$.
Nevertheless, a fit for $\Omega$ using the DDK model yields
$\Omega\, = \, (320\, \pm \, 104)^\circ$. While
consistent with the measurement of $\Omega$ in Paper II,
this is not very restrictive, i.e., we see no sign
of a constraint on $\Omega$ introduced by the annual orbital parallax
beyond the constraints already introduced by the measurement of
$\dot{x}$.

This is not surprising: this is a very small effect that
has only been detected for pulsars with extremely
high timing precision like PSRs J0437$-$4715, J1713+0747 and J1909$-$3744 
\citep{Reardon2016,Fonseca2016,Desvignes2016}.
However, given its combination of small distance and relatively large orbital
size, PSR~J2222$-$0137 is a candidate for the detection of this effect,
particularly if can be observed with more
sensitive telescopes. Since Arecibo is not an option
(the pulsar is just below that telescope's Southern limit)
this will only be possible with FAST \citep{Smits2009} or the SKA \citep{ShaoSKA}.

Since the DDK model introduces new orbital effects, it is
important to verify whether they have an impact on other orbital
parameters. The answer is no: the Shapiro delay
masses obtained with this model are 0.15 $\sigma$
smaller and the $\dot{\omega}_{\rm obs}$ is 0.2 $\sigma$ smaller
compared to the values from the DD model, whether we assume
the VLBI $\Omega$ or not. The astrometry
is unchanged. As the timing precision
improves, it will be essential to do this correction, and for that
the VLBI measurements are very important.

\subsubsection{Contribution to the observed periastron advance}

Since the observed $\dot{\omega}$ appears to be slightly larger than
the GR expectation, it is important to know whether there are
extra contributions to this effect that have not been taken into account.

Given the large proper motion of the system, there will be a kinematic
contribution to $\dot{\omega}$ caused by the changing viewing geometry, this
is given by:
\begin{equation}
  \dot{\omega}_{\rm K} =  \frac{\mu_{\rm T}}{\sin i} \cos(\Theta_{\mu} - \Omega), \label{eq:OmegaK}
\end{equation}
where all quantities are as in the equation above. Thus $\dot{\omega}_{\rm K}\,=\, 
-0.5^{+3.3}_{-4.3} \times 10^{-6\, \circ} \, \rm  yr^{-1}$. Even if we had no constraints 
whatsoever on $\Omega$, we would still have $|\dot{\omega}_{\rm K}|\, \leq \, 1.25 \times 
10^{-5 \, \circ} \rm\,  yr^{-1}$. This is more than two orders of magnitude smaller than 
the measurement uncertainty of $\dot{\omega}_{\rm obs}$, $2.9 \times 10^{-3\, \circ} 
{\rm yr^{-1}}$. We thus conclude that this is not the reason for the larger than
expected value of $\dot{\omega}$.

%==========================================================================================

\section{PSR J2222$-$0137 as a gravitational wave laboratory}
\label{sec:GWs}

The large difference in compactness between a NS and a WD makes pulsar-WD binaries ideal
laboratories for gravity theories that violate the strong equivalence principle 
\citep{ShaoWex2016}. In such alternatives to GR, the leading contribution in the loss of 
orbital energy by gravitational waves, generally, comes from dipolar radiation. The 
radiation reaction of dipolar waves enters the equations of motion of a binary system 
already at the 1.5 post-Newtonian level (i.e.\ order $v^3/c^3$), and is therefore expected 
to dominate the damping of the orbit in systems like PSR~J2222$-$0137. For the 
scalar-tensor theories, like the ones studied in \cite{DE92,DE93}, one finds for the change 
in the orbital period by dipolar radiation damping
\begin{equation}
  \dot{P}_b^{\rm D} = -2\pi\frac{G_\ast}{c^3}\,
     \frac{m_p m_c}{m_p + m_c} \, \left(\frac{P_b}{2\pi}\right)^{-1}
     \frac{1+e^2/2}{(1 - e^2)^{5/2}} \, (\alpha_p - \alpha_c)^2 \;,
\label{eq:PbdotD}
\end{equation}
where $G_\ast$ denotes the bare gravitational constant, and $\alpha_p$ and $\alpha_c$ are
the effective scalar couplings of pulsar and companion respectively. In the discussion here,
for simplicity, we will mostly refer to the mono-scalar-tensor theory investigated in
detail in \cite{DE93,DE96}. Nevertheless, the (generic) limits we present below apply 
to a wider class of scalar-tensor theories, and to some extent to a large range of gravity
theories with dipolar radiation. The quantities $\alpha_p$ and $\alpha_c$ in equation~
(\ref{eq:PbdotD}) have a (non-linear) dependence on the bodies' gravitational binding 
energy. For the weakly self-gravitating WD companion the effective scalar coupling can be
approximated by the linear matter-scalar coupling constant, i.e.\ $\alpha_c \approx 
\alpha_0$. In the meantime, a Solar System experiment \citep{BIT2003}
constrains $|\alpha_0| \lesssim 0.003$ 
\citep{Will2014}. The pulsar's effective scalar coupling $\alpha_p$, on the other hand, may 
be of order one, even for a vanishing  $\alpha_0$ \citep{DE93}. Given the uncertainties in 
the masses of PSR~J2222$-$0137, we can safely approximate $G_\ast$ by Newton's 
gravitational constant $G \equiv G_\ast(1 + \alpha_0^2)$, and use ${\rm T}_\odot$ instead 
of $G_\ast {\rm M}_\odot/c^3$ in equation~(\ref{eq:PbdotD}). Inserting the numbers from 
Table~\ref{table:timing} into equation~(\ref{eq:PbdotD}) gives
\begin{equation}
  \dot{P}_b^{\rm D} = -(697 \pm 19) \times 10^{-12} \, (\alpha_p - \alpha_0)^2 \,.
\label{eq:PbdotDNum}
\end{equation}
In case of a strongly scalarized PSR~J2222$-$0137 ($\alpha_p \sim 1$) 
equation~(\ref{eq:PbdotDNum}) gives an orbital decay that is in conflict with the derived
intrinsic change of the orbital period ($\dot{P}_{b, \rm int}$ in 
table~\ref{table:timing}) by about four orders of magnitude. In turn, this implies a tight 
upper limit on the scalar charge of PSR~J2222$-$0137. A detailed analysis, which properly 
accounts for all the uncertainties and correlations through Monte Carlo simulations, gives 
the following (generic) limit
\begin{equation}
  |\alpha_p| < 0.02 \quad \mbox{(95\% C.L.)} \,.
\label{eq:alphapLimit}
\end{equation}
Since the masses for the PSR~J2222$-$0137 are derived from the Shapiro delay, caused by the
(weakly self-gravitating) WD companion, one also has to account for a modification of the
gravitational constant in the mass function (see equation~5.8 in \citealt{DE96}). However, 
this strong-field modification is sub-leading in our test, since in the PSR~J2222$-$0137
system the effective gravitational constant is given by $G_\ast(1 + \alpha_p\alpha_0)$,
and $|\alpha_0| \ll 1$.

The limit from equation~(\ref{eq:alphapLimit}) is an order of magnitude weaker than the limit derived for 
PSR~J1738+0333 ($1.46\,\Msun$) in \cite{Freire+12} and still a factor of four weaker than the 
limit derived by \cite{Antoniadis} for PSR~J0348+0432 ($2.0\,\Msun$). However as first 
pointed  out by \cite{Shibata+2014}, depending on the EoS, a $\sim 1.8\,\Msun$ could still be 
strongly scalarized, in spite of the tight limits at 1.46 and 2.0\,$\Msun$. Because of 
this, the limit given above has its own importance. In fact, in a
recent analysis, \cite{Shao2017} have used a set of pulsars with different masses, in order to place the 
best constraints on such strong field scalarization, for a wide range of EoS. Such 
constraints, as discussed by these authors, are of great importance for LIGO, Virgo and 
future gravitational wave observatories. The limit of PSR~J2222$-$0137 makes, because of 
the pulsar's mass, an important contribution to the constraints of \cite{Shao2017}. 

The precision of this test will improve greatly in the future:
First, because $m_p$ and $m_c$ will be much better known; this is important for the
interpretation of the experiment, in particular for a precise calculation of the 
expected $\dot{P}_b$ from dipolar GWs\footnote{The uncertainty on $\dot{P}_b$ from
the quadrupolar GR contribution $\dot{P}_{b, \rm GR}$ is already so small that
we can ignore it; improvements in the mass measurements will further reduce the
uncertainty of this quantity in the future.}:
For some equations of state (and some scalar-tensor theories of gravity),
the dipolar term varies extremely rapidly with $m_p$,
so its precise measurement is very important. Second, because, as mentioned above,
the uncertainty of $\dot{P}_b$ ($\delta \dot{P}_b$),
decreases with $T^{-5/2}$, so a factor of $\sim 10$ improvement 
(i.e, $\delta\dot{P}_{b, \rm obs} \, = \, 0.008 \, \times \, 10^{-12}\, \rm s\, s^{-1}$)
will be achieved  when our timing baseline is 10 years long (i.e., within 6 years from now).
A factor of 50 improvement over the current limit
($\delta\dot{P}_{b, \rm obs}\, = \, 0.0016 \, \times \, 10^{-12}\, \rm s\, s^{-1}$)
would require a total timing baseline of 19 years, i.e., it will be achieved in about 15 years.
This can be significantly reduced if a much more sensitive radio telescope
is used to time the pulsar, like the SKA \citep{ShaoSKA}.
At that time the precision on $\dot{P}_{b, \rm obs}$ will be similar to the current uncertainty
from kinematic contributions (see Subsection~\ref{sec:Pbdot})
and we will have a limit on $\alpha_p$
for this pulsar similar to that derived for PSR~J1738+0333. 
However, the better models of the Galactic potential that will result from the
GAIA mission and the much improved distance measurements that will be
made possible with the SKA should allow a significant improvement in the estimate of
the kinematic terms, which will translate directly into an improvement of the
precision of this experiment. 

%------------------------------------------------------------------------------------------

\section{Modelling the formation of PSR~J2222$-$0137}
\label{sec:modelling}

Despite the relatively massive compact objects in this binary, the 
formation of PSR~J2222$-$0137 is expected to resemble the standard scenario for
forming binaries with recycled pulsars and massive WDs (e.g. \citealt{TLK12}).
The zero-age main sequence (ZAMS) progenitor system of PSR~J2222$-$0137 may
have contained two ZAMS stars with a primary mass of $20 - 25\,\Msun$ and a
secondary mass of 6 -- 8 $\Msun$ (see discussion below). Assuming this binary
formed a common envelope (CE, \citealt{Ivanova13}) once the
primary star became a giant star and initated mass transfer, the secondary star
was unable to accrete much material and its mass may have remained roughly
constant until the supernova (SN) explosion of the naked core of the primary
star. After the formation of the NS, the system briefly becomes an
intermediate-mass X-ray binary (IMXB) when its relatively massive companion star
fills its Roche lobe and starts transferring mass to the NS (via Roche-lobe overflow, or RLO).
As a consequence of the large mass-ratio ($q\simeq 3-4$)
between the donor star and the accreting NS, the mass transfer becomes
dynamically unstable and leads to the formation of a second CE in which the
hydrogen-rich layer is removed from the companion star and a naked helium star is formed.

The further evolution of such NS-helium star binaries leads to a final phase of
mass transfer (so-called Case~BB RLO) once the helium star becomes a giant star,
e.g. \cite{Habets86}, \cite{Dewi02}, \cite{Ivanova03}, \cite{TLK12}, \cite{Lazarus14}, and
\cite{Tauris15}.   Here we follow the method of detailed
modelling outlined in \cite{Lazarus14}.

For the progenitor of the massive $1.293 \, \pm \, 0.025\, \Msun$ ONeMg~WD we find that
the mass of the helium star was $2.4-2.5\, \Msun$. The orbital period prior to
Case~BB RLO was about $1.2\,$-$\,1.3\;{\rm days}$  and the total duration of this
mass-transfer phase was only $\le 20\,000\;{\rm yr}$. This rapid phase of
recycling combined with the current spin period of 32~ms places some interesting 
constraints on the accretion efficiency of the NS. Before illuminating the
accretion history, we first discuss the WD cooling age and the spin history of
this pulsar. 

%------------------------------------------------------------------------------------------

\subsection{Spin-down history of the pulsar and cooling age of the companion}

Following Paper~I, we use the WD cooling tracks of 
\cite{Bergeron11} to estimate its age. The most massive WD listed 
in their tables\footnote{
http://www.astro.umontreal.ca/$\sim$bergeron/CoolingModels/}  has a mass of $1.30\, \Msun$, 
consistent with our estimate for the companion of PSR~J2222$-$0137. This massive WD cools 
down to a surface temperature below 3000~K in 3.4~to~4.7~Gyr,  depending on the hydrogen 
content in the outer layers. Given the non-detection of the companion of PSR~J2222$-$0137, 
however, the end of the recycling phase could have occurred a longer time ago -- in 
principle, between 3.4 and 13~Gyr. We note that the larger mass estimate of the WD implies 
that the system is not necessarily as old as implied in Paper I. For a $1.05\,\Msun$ CO~WD
(the WD mass estimate from Paper~I),
the cooling tracks of \cite{Bergeron11} reach $T_{\rm eff}=3000\,{\rm K}$ in 7.4~to~9.7~Gyr.
Therefore, a $1.30\,\Msun$ ONeMg~WD cools to this temperature 4--5~Gyr faster than
a $1.05\,\Msun$ WD (which probably has a CO composition interior instead, although this
difference in chemical composition should not affect the cooling timescale by much).

From the observed spin period (32.8~ms) and period derivative ($\dot{P}_{\rm
int}=1.75 \times 10^{-20}$), we can probe the spin-down history of
PSR~J2222$-$0137 given the above cooling age constraint from the ONeMg~WD (cf.
Section~5.1.1 in Lazarus~et~al.~2014). Assuming a standard spin evolution with a
braking index of $n=3$, we find an initial $P$ (after recycling) of 30~ms
and 25~ms, for an age of 3.4~Gyr and 12.4~Gyr, respectively. The limited slow
down of the spin period, even on a Hubble~time, is caused by the relatively small
B-field of this pulsar; it implies that after recycling the pulsar had a
spin period similar to what is has today. This is useful when probing the
accretion history of the system.

%------------------------------------------------------------------------------------------

\subsection{Accretion history and NS birth mass of PSR~J2222$-$0137}

If the accretion rate onto the NS is limited by the Eddington luminosity, the NS
is only able to accrete $7-8\times 10^{-4}\, \Msun$ during the short-lasting
Case~BB RLO. This amount is a factor of $\sim\!4$ smaller than what is needed to
reach a recycled spin period of 30~ms according to Eq.(14) in \cite{TLK12}.
Hence, we can conclude that either PSR~J2222$-$0137 accreted at a
slightly super-Eddington rate (as also found for PSR~J1952+2630 in
Lazarus~et~al.~2014), or that the geometric factor $f(\alpha,\xi,\phi,\omega_c)$
introduced in Eq.~(13) in \citet{TLK12} should be less than unity. 

Whereas some MSPs must evidently have been inefficient accretors
\citep[e.g.][]{avk+12}, 
it has recently been argued that NS accretors in the ultraluminous X-ray binaries
M82~X-2 \citep{Bachetti14} and ULX-1 in NGC~5907 \citep{ibs+16} may even accrete at a rate of up to ~100 times the
Eddington limit to explain the observed X-ray luminosities of these sources ($L_X\sim
10^{40}\;{\rm erg\,s}^{-1}$) and the high spin-up rate of the latter NS. 
However, these sources are suggested to harbor very strong B-field NSs and are most likely not
representative of pulsars in general.
Nevertheless, we conclude that observations of accreting NSs and
recycled MSPs continue to challenge the details of physical models for the
accretion of matter onto  magnetized compact objects. 

Regardless of the abovementioned uncertainty in the exact amount of material
accreted by  PSR~J2222$-$0137, the total amount of matter accreted was probably
at most a factor of a few $10^{-3}\,\Msun$, or less than $10^{-2}\,\Msun$ including wind accretion prior to the onset of RLO/CE of the IMXB progenitor system. Hence, the NS birth mass of
PSR~J2222$-$0137 was almost identical to the presently measured mass
of $1.76 \, \pm\, 0.06\, \Msun$. This is the largest
birth mass inferred for any radio pulsar. Of the other known $2\,\Msun$
pulsars, J0348+0432  could (but need not) potentially have accreted several $0.10\, \Msun$
\citep{Antoniadis}, and  PSR~J1614$-$2230 was the previous record holder
with a minimum estimated birth mass of about $1.7\, \Msun$ \citep{TLK11,Lin11}.
A few NSs in high-mass X-ray binaries are likely born with larger masses \citep[e.g.][]{tkf+17}.

A NS birth mass of close to $2\,\Msun$ may, from a stellar evolution point of view,
suggest a relatively massive progenitor star of at least
$20\,\Msun$ \citep[see discussions in][and references therein]{TLK11}, although
many aspects of the final outcomes of massive star evolution are still uncertain \citep{Langer}. More
importantly, such large NS birth masses are not easily obtained from current SN explosion
modelling \citep{Ugliano12,PT15,ejw+16,sew+16} -- even when fallback is included in 1D simulations.
We notice that the parametrized modelling of \citet{mhlc16} results in some NSs with masses up to $2\,\Msun$. Interestingly enough, they originate from progenitor stars with initial ZAMS masses of only $\sim\!15\,\pm 1\,\Msun$,
while less massive NSs (as well as black holes) are produced for initial progenitor star masses up to $\sim\!27\,\Msun$.
The lack of massive NSs from more massive progenitor stars ($20-25\,\Msun$)
in the work of \citet{mhlc16} is possibly caused by their assumed maximum possible NS mass 
of $2.05\,\Msun$. Their results depend on the amount of accretion after shock revival and 
need to be confirmed by 3D modelling. Furthermore, it is possible that magnetohydrodynamic 
effects could still allow for explosions in cases where the neutrino-driven mechanism 
fails (H.-T. Janka, 2017, priv.~comm.). Finally, it should be noted that above-mentioned SN 
simulations are based on different sets of progenitor stars, e.g. evolved with different 
wind-mass loss assumptions, and only considered explosions of isolated (i.e. non-stripped) 
stars.

Although the birth mass of PSR~J2222$-$0137 is relatively large compared to that of 
other radio pulsars, it is still in accordance with, although a bit on the high side of, 
the NS birth mass spectrum of current SN modelling 
\citep[see the results in][]{Ugliano12,PT15,ejw+16,sew+16,mhlc16}.  

%=========================================================================================

\section{Conclusions}
\label{sec:conclusions}

In this work we present the results of four years
of timing of PSR~J2222$-$0137.
The parallax we measure matches the VLBI
parallax well within 1 $\sigma$, this is encouraging since this quantity
is especially sensitive to systematic effects in the timing.
The proper motions match  as well, but not so closely.
We find a significant and very robust offset of our timing position
and the VLBI position, which we have investigated in detail.
This offset is not caused by DM variations caused by the Sun or otherwise,
and it does not change for different choices of Solar System ephemeris.
Based on recent VLBI studies, we conclude that the uncertainty 
of the absolute VLBI position measurement for
PSR~J2222$-$0137 was likely under-estimated; however, it is also not
clear whether this is the explanation for the observed offset.

We have obtained masses for both PSR~J2222$-$0137 and its companion
that are significantly larger than the masses presented in Paper~I.
The reasons for this
are not entirely clear, but we believe our results to be more accurate because
a) we have a larger number of observations and much improved orbital coverage
and a larger baseline, b)
we measure the rate of advance of periastron $\dot{\omega}$ with high
significance, and (assuming GR) the observed value is in strong disagreement
with the lower mass values presented in Paper~I, but in a marginal agreement with ours.
We conclude therefore that the measurement presented here is more
accurate. This is corroborated by the observation in Paper~I that
the mass values from Shapiro delay change significantly when they fit
for $\dot{\omega}$, indicating a correlation between these parameters.
The highly significant measurement of $\dot{\omega}$ presented in this work
is therefore important to reduce parameter correlations and allow the
determination of accurate masses.
Furthermore, based on the measurement of $\Omega$ presented in Paper I,
we were able to determine the inclination of the system from the measurement of
$\dot{x}$.
Continued timing will achieve a high precision measurement of the component masses and
inclination because the uncertainties in the measurement of $\dot{\omega}$
and $\dot{x}$ decrease with the timing baseline $T$ as $T^{-3/2}$.

The mass measurement for PSR~J2222$-$0137 and the precise knowledge of its distance
make its system interesting for tests of gravity theories, in particular for alternatives
to GR that violate the strong equivalence principle and predict the existence of dipolar
gravitational waves. The mass of PSR~J2222$-$0137 falls into a range that so far is 
poorly constrained in terms of phenomena like spontaneous scalarization \citep{ShaoWex2016},
and for this reason makes an important contribution in testing scalar-tensor theories of
gravity. Among others, this is important for future gravitational wave observations
with ground-based gravitational wave detectors, like LIGO and Virgo \citep{Shao2017}.

The improved mass measurements are also important for understanding the evolution
of the system. With a mass of $1.293\, \pm\, 0.025\, \Msun$; the companion
of PSR~J2222$-$0137 is one of the most massive WD companions to any pulsar known
(e.g., \citealt{Ozel&Freire2016}).
Such a massive WD is expected to crystalize and reach
surface temperatures below 3000~K on a total timescale of 3~to~4~Gyr, explaining
the non-detection of this WD in the most sensitive optical observations made to
date. Detecting this WD with more sensitive optical observations, measuring its
color and estimating its radius represent high priority goals.

Furthermore, given the relatively short timescale for the evolution of the progenitor
star of the WD, 
the accretion episode could not have been long, and the total amount of mass
transferred was likely $<\, 10^{-2} \, \Msun$. Therefore this pulsar was born with a
mass that was similar to its current mass, making it the largest NS birth mass
measured to date for any radio pulsar. This measurement shows, furthermore, that the pulsars
in IMBPs also have a wide range of masses: from $1.24\, \pm \, 0.11 \, \Msun$ for PSR~J1802$-$2124 
to $1.76\,\pm 0.06\,\Msun$ for PSR~J2222$-$0137 and even $1.928\,\pm\,0.017\,\Msun$
if we include PSR~J1614$-$2230 in this class \citep{Fonseca2016}.

The mass measurement presented in this work plus
those of PSRs~J1614$-$2230 \citep{Fonseca2016}, J0348+0432 
\citep{Antoniadis}, J1903+0327 \citep{FreireJ1903} and J1946+3417 \citep{Barr2017}
show that massive NSs are not rare in the Universe.
This becomes even clearer if we take into account the fact that we find this
NS in one of the two known binary pulsars with measured masses within 300~pc of the Earth.

%=========================================================================================

\acknowledgments
We thank Thomas Janka for discussions on NS birth masses, Adam Deller for
discussions on the VLBI astrometry and Alessandro Ridolfi and John Antoniadis for proofreading
and useful suggestions.
The Nan\c{c}ay Radio Observatory is operated by the Paris Observatory, associated with
the French Centre National de la Recherche Scientifique (CNRS). We acknowledge financial
support from ``Programme National de Cosmologie and Galaxies'' (PNCG) of CNRS/INSU, France.
PCCF gratefully acknowledges financial support by the European Research Council for the
ERC Starting grant BEACON under contract No. 279702 during most of his work on this project.
MK and GD acknowledge the financial support by the European Research Council for the ERC Synergy Grant BlackHoleCam under contract no. 610058. 
%=========================================================================================

\facilities{Effelsberg 100-m Radio Telescope, Nan\c{c}ay Radio Telescope, Lovell 76-m radio telescope} 
\software{Numpy, PSRCHIVE, tempo}

%=========================================================================================

\bibliographystyle{yahapj}
\bibliography{ref}

\acknowledgments

\end{document}